\documentclass{article}

\usepackage{color}
\usepackage{graphicx}
\usepackage{latexsym}
\usepackage[algoruled,vlined,algosection]{algorithm2e}
\SetKwInOut{Input}{Input}
\SetKwInOut{Output}{Output}

\newtheorem{thm}{Theorem}
\newtheorem{lemma}[thm]{Lemma}

\newtheorem{proposition}[thm]{Proposition}

\newcommand{\DSKL}{{\sc Directed Spanning $k$-Leaf}}

\newcommand{\LL}{\mbox{\sc Leaf}}
\newcommand{\BO}{\mbox{\sc BrSucc}}
\newcommand{\Hd}{\mbox{\sc Head}}
\newcommand{\T}{\mbox{\sc Tail}}
\newcommand{\Back}{B}
\newcommand{\HdBack}{\mbox{\sc HB}}

\newcommand{\bs}{\backslash}

\newcommand{\proof}{{\bf Proof: }}
\newcommand{\QED}{\mbox{}\hspace*{\fill}{$\Box$}\medskip}

\title{An FPT Algorithm for Directed Spanning $k$-Leaf}
\author{Paul Bonsma\footnote{Supported by the Graduate School ``Methods for Discrete Structures'' in Berlin, DFG grant GRK 1408.}\\
\small Technische Universit\"{a}t Berlin,\\
\small Institut f\"{u}r Mathematik, Sekr. MA 6-1,\\ 
\small Stra\ss{}e des 17. Juni 136, 10623 Berlin, Germany\\
\small bonsma@math.tu-berlin.de\\
\\
Frederic Dorn\\
\small Humboldt-Universit\"{a}t zu Berlin\\
\small Institut f\"{u}r Informatik\\
\small Unter den Linden 6, 10099 Berlin, Germany\\
\small dorn@informatik.hu-berlin.de
}

\date{November 26, 2007}

\begin{document}

\maketitle

\begin{abstract}
An out-branching of a directed graph is a rooted spanning tree with all 
arcs directed outwards from the root. We consider the problem 
of deciding whether a given digraph $D$ has an out-branching 
with at least $k$ leaves (\DSKL). 
We prove that this problem is fixed parameter tractable, when $k$ is chosen as the parameter. Previously this was only known for restricted classes of directed graphs. 

The main new ingredient in our approach is a lemma that shows that given a locally optimal out-branching of a directed graph in which every arc is part of at least one out-branching, either an out-branching with at least $k$ leaves exists, or a path decomposition with width $O(k^3)$ can be found. This enables a dynamic programming based algorithm of running time $2^{O(k^3 \log k)} \cdot n^{O(1)}$, where $n=|V(D)|$.
\end{abstract}

\section{Introduction}

Directed graphs or \emph{digraphs} are graphs with vertices connected by oriented \emph{arcs}. 
Directed graph problems are in general harder to solve than their analog on undirected graphs, since undirected graphs may be seen as a special case of digraphs, namely directed graphs with arcs in both directions.
Considering $NP$-hard problems in particular, the results on algorithms for problems on undirected graphs are many, compared to the corresponding versions on digraphs.
Though there exist problems where the fastest algorithm works for both types of graphs---as a prominent example take the \emph{colour coding} algorithm for {\sc Longest Path} of \cite{AlonYZ95}---the algorithmic ideas used in many fast solutions for problems on undirected graphs do not apply to the directed case.
 
We consider the $NP$-hard problem of finding spanning trees with maximum number of leaves ({\sc Maximum Leaf Spanning Tree}). 
This is a well-studied problem on undirected graphs, see e.g.~\cite{BonsmaBW03,bonsmazickfeld, DingJS01, FellowsMRS00, FominGK06, GalbiatiMM97, KW91, LuR98, Solis-Oba98} for a selection of approximation algorithms, exact algorithms and results from extremal graph theory. Note that the problem is closely related to connected dominating set; instead of maximizing the number of leaves we may also choose to minimize the number of non-leaves, which form a connected dominating set.

One reason that these problems 
are well-studied is because of their ample applications, for instance in (wireless) networks. From this practical viewpoint, {\sc Maximum Leaf Spanning Tree} is even more interesting on digraphs, though theoretically only little studied. In one typical application, locations of transmitter nodes are given, and every node can transmit information to nearby nodes. The goal is to select routing nodes plus a root, such that messages from the root can be relayed through the routing nodes to every other node (broadcasting), or the other way around (e.g. sensor networks). For cost considerations, the number of routing nodes should be minimized, or equivalently, the number of {\em leaf nodes} should be maximized. When we assume that transmission capabilities of nodes are uniform, this problem is modeled by {\sc Maximum Leaf Spanning Tree} on undirected graphs. This assumption is however not always justified~\cite{HuangH89,Liang02,TWL07}, which leads to the formulation of the problem we consider.

For digraphs we will use notions that are defined for undirected graphs, such as paths, cycles, trees, connectedness, and vertex neighborhoods. These are defined as expected, where arc directions are irrelevant. An {\em out-tree} of a digraph is a subgraph that is a rooted tree, where all vertices have in-degree 1 except for one which has in-degree 0, the {\em root}. An {\em out-branching} of a digraph is a {\em spanning} out-tree. A {\em leaf} of a digraph is a vertex with out-degree 0. The problem is now defined as follows.\\

\noindent
{\sc Directed Spanning $k$-Leaf}:\\
{\sc instance}: A connected digraph $D$ and an integer $k$.\\
{\sc question}: Does $D$ contain an out-branching with at least $k$ leaves?\\

We consider the \emph{parameterized} version of this problem, where we choose $k$ as the parameter. We are interested in \emph{fixed parameter tractable (FPT)} algorithms, which are algorithms with a time complexity of the form $f(k) \cdot n^{O(1)}$, where $f(k)$ is a function only depending on $k$, the {\em parameter function}, and $n=V(D)$.

After many improvements (see e.g.~\cite{FellowsMRS00,BonsmaBW03}), the current fastest FPT algorithm for undirected graphs appears in~\cite{bonsmazickfeld}, with a running time of $O^*(6.75^k)+O(m)$, with $m=|A(D)|$ being the number of arcs. This problem is an example where one of the essential ideas for the undirected case does not hold for digraphs, namely that \emph{any} tree with $k$ leaves can be extended to a spanning tree with at least $k$ leaves. See below for an example.

Tackling an open problem posed by Michael Fellows in 2005~\cite{Cesati}, Alon et al.~\cite{AlonFGKS07} were the first to prove that {\sc Directed Spanning $k$-Leaf} admits an FPT algorithm when restricted to a certain graph class that includes for instance strongly connected digraphs and acyclic digraphs, with parameter function $f(k)= 2^{O(k^2 \log k)}$. In~\cite{AlonFGKS07II}, the same authors improve that function to $2^{O(k \log^2k)}$ for strongly connected graphs, and $2^{O(k\log k)}$ for acyclic graphs.
The question whether the problem admits an FPT algorithm for all digraphs remained open, and was posed again in~\cite{AlonFGKS07II,AlonFGKS07,DemaineGMS07}.

In their approach, Alon et al.\ consider classes of graphs where the maximum number of leaves is the same for out-trees and out-branchings (or where no out-branching exists, which is the trivial case). Hence, instead of creating an out-branching with at least $k$ leaves, it suffices to find an out-tree with at least $k$ leaves. 
They show that, given a locally optimal out-branching that has less than $k$ leaves, either such an out-tree can be found, or a path decomposition can be given with width bounded by a function of $k$. This allows standard dynamic programming approaches to be used.
We sketch and interpret the main idea of the proof of this statement. The authors decompose the tree into directed paths, and consider the number of \emph{backward} arcs for a combination of such a path $P$ and vertex $v\in V(P)$ on this path---loosely speaking, those are sets of arcs that are not part of the tree and form a directed cycle with a part of $P$, that contains $v$.
It is then shown that either a path and vertex exist for which the number of backward arcs is at least $k$, which immediately gives an out-tree with $k$ leaves (rooted at $v$), or for each path and each vertex on this path it is less than $k$, which can be used to find a path decomposition with width bounded by a function of $k$. For this last step, the local optimality of the out-branching is essential.

\textbf{Our contribution.} 
In this paper we answer the above question positively, by providing the first FPT algorithm for \DSKL\ that works for all digraphs. An overview of our algorithm is given in Section~\ref{sect:overview}. It uses the same general approach as introduced in~\cite{AlonFGKS07II,AlonFGKS07}: we start with a locally optimal tree, and use it to either find an out-branching with at least $k$ leaves (when the number of backward arcs is large), or to find a path decomposition of width bounded by a function of $k$ (when this number is small). 

\begin{figure}[h]
\centering
$\input{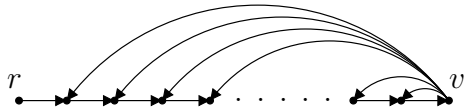}$
\caption{A graph with a leafy out-tree but no leafy out-branching.}
\label{fig:useless}
\end{figure}

However an addition to this approach is needed, since it can not work for all digraphs: consider Figure~\ref{fig:useless}~\footnote{Example given by Gregory Gutin at a lecture at the Fall School on Algorithmic Graph Structure Theory, Blankensee 2007}. 
This digraph has an out-tree with $n-2$ leaves, rooted at $v$, but the unique out-branching, which is rooted at $r$, has only one leaf. More importantly, this example shows that the ratio between the maximum number of leaves of an out-branching on one hand, and the number of backward arcs or the pathwidth on the other hand may be arbitrarily bad.
But if one takes a closer look at the arcs of the out-tree, one may observe that they are irrelevant for the problem we consider; they do not appear in any out-branching. 
We will first remove all such arcs, which are called {\em useless arcs}, see Section~\ref{sect:uselessarcs}. For the remaining graph, we consider a locally optimal out-branching, and deduce its properties in Section~\ref{sect:oneoptoutb}.
However, this still does not enable us to apply the ideas from the former algorithms.

Therefore, in Section~\ref{sect:mainlemma}, we prove a key lemma which is our main new contribution: we show that, in a graph without useless arcs for which a locally optimal out-branching with less than $k$ leaves is given, if there is any point where the number of backward arcs is at least $6k^2$, an out-branching with at least $k$ leaves can be found. If not, then 
we create a path decomposition of width bounded by $6k^3$ in Section~\ref{sect:pathdecomp}. Together with a dynamic programming procedure, these ingredients give the FPT algorithm, which is summarized in Section~\ref{sect:algsummary}. 
The relatively short proof of our main lemma is made possible by making heavy use of the partial order structure defined by an out-branching, which is a useful new approach for this problem.
We first start in Section~\ref{sect:prelim} with definitions.
  
\section{Preliminaries}
\label{sect:prelim}

\paragraph{General definitions}

For basic graph theoretic definitions see~\cite{Diestel}, and for directed graphs in particular see~\cite{BangJG00}.
For a digraph $D$, $V(D)$ denotes the set of vertices and $A(D)$ the set of arcs. Arcs are 2-tuples $(u,v)$ where $u\in V(D)$ is called the {\em tail} and $v\in V(D)$ the {\em head}.
For an arc set $B$, $\Hd(B)$ is the set of heads of arcs in $B$, 
$\T(B)$ is the set of tails.
The underlying undirected graph of a digraph $D$ is denoted by $D_u$.
A~\emph{dipath} is a graph with vertex set $\{v_1,v_2,\ldots,v_r\}$ and arc set
$\{(v_i,v_{i+1}):i=1,\ldots,r-1\}$. This will also be called a {\em $(v_1,v_r)$-dipath}. With such a dipath we associate an order from $v_1$ to $v_r$, for instance when talking about the first arc of the path that satisfies some property.

We define now a \emph{path decomposition} $(X_1,\ldots,X_q)$ of an (undirected) graph $G$ 
as a collection of subsets $X_i$
of $V(G)$ (\emph{bags}) such that
\begin{enumerate}
\item
$\bigcup_{i =1,\ldots,q} X_{i}
= V(G)$, 
\item
for each edge $vw \in E(G)$, there exists an $i\in
\{1,\ldots,q\}$ such that $v,w\in X_{i}$, and
\item
for each $v\in V(G)$,  there exist $i,k$ with $1\leq i \leq k \leq q$ such that $v \in X_j$ for all $j$ with $i \leq j \leq k$ and $v \notin X_{\ell}$ for all $\ell$ with $\ell < i$ or $\ell > k$ .
\end{enumerate}
The \emph{width} of a path decomposition $(X_1,\ldots,X_q)$
equals $\max_{i \in \{1,\ldots,q\}} \{|X_{i}| - 1\}$.

A {\em partial order} is a binary relation that is reflexive, antisymmetric and transitive. A {\em strict partial order} is irreflexive and transitive. Partial orders will be denoted by $\preceq$, and strict partial orders by $\prec$. If for every pair of elements $u$ and $v$, either $u\preceq v$ or $v\preceq u$ holds, the order is a {\em linear order}. For strict partial orders, the corresponding notion is a {\em strict linear order}.

\paragraph{Definitions for out-trees and out-branchings}

A subtree $T$ of a digraph $D$ is an \emph{out-tree} if it has only one vertex of
in-degree zero, its \emph{root}. If $T$ is a spanning out-tree of $D$, i.e. $V(T)=V(D)$, then we call $T$  an \emph{out-branching} of $D$. 
The vertices of $T$ of out-degree zero are \emph{leaves} and the vertices of out-degree at least two are called \emph{branch vertices}.
Let $\LL(T)$ denote the set of leaves of $T$, and
$\BO(T)$ the vertices of $T$ that have a branch vertex of $T$ as in-neighbor. 
Note that $\LL(T) \cap \BO(T)$ may not be empty.
 
If there exists a dipath in $D$ from vertex $u$ to vertex $v$, we say $v$ is {\em reachable} from $u$ (within $D$). The set of all vertices that are reachable from $u$ within $D$ is denoted by $R_D(u)$. (This set includes $u$ itself.)
Let $T$ be an out-tree. Then we write $u\preceq_T v$ if $v\in R_T(u)$, and $u\prec_T v$ if in addition $v\not=u$.
\begin{proposition}
Let $T$ be an out-tree. The relation $\preceq_T$ is a partial order on $V(T)$.
\end{proposition}
\begin{proposition}
\label{propo:extendingouttree}
Let $T$ be an out-tree of a digraph $D$, with root $r$. Then $D$ has an out-branching $T'$ with root $r$, that contains $T$, if and only if $R_D(r)=V(D)$.
\end{proposition}
\proof
First observe that if there exists any out-branching with root $r$ then it is obvious that $R_D(r)=V(D)$. To prove the other direction, suppose that for the root $r$ of $T$, $R_D(r)=V(D)$ holds. Consider a maximal out-tree $T'$ with root $r$ that contains $T$. Suppose $V(T')\not=V(D)$. Let $u\in V(D)\bs V(T')$. Since $R_D(r)=V(D)$, there exists a $(r,u)$-dipath. Consider the first arc on this path with head not in $V(T')$: this arc can be added to $T'$, a contradiction with the maximality of $T'$.
\QED

Let $T$ be an out-branching of $D$, and let $(u,v)\in A(D)\bs A(T)$, where $v$ is not the root of $T$. The {\em 1-change for $(u,v)$} is the operation that yields $T+(u,v)-(w,v)$, where $w$ is the unique in-neighbor of $v$ in $T$.
We call an out-branching $T$ \emph{1-optimal} if there is no 1-change for an arc of $A(D) \bs A(T)$ that results in an out-branching $T'$ with more leaves. 

Let $D$ be a digraph with a vertex $r$ such that $R_D(r)=V(D)$. An arc $(u,v)$ of $D$ is {\em useless for $r$} if there is no out-branching of $D$ that contains $(u,v)$ and has $r$ as root.\\

\begin{figure}[h]
\centering
$\input{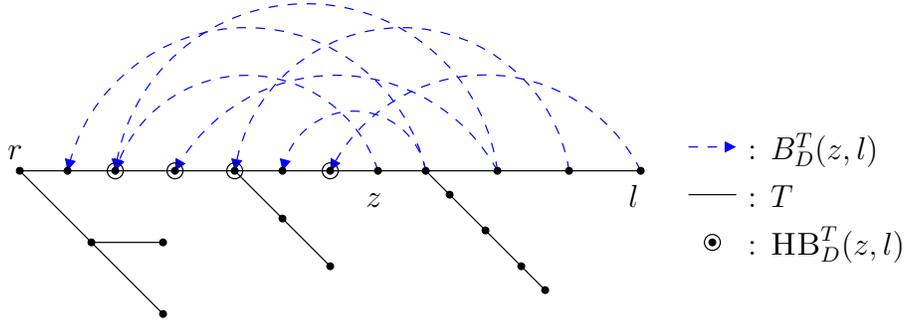}$
\caption{The sets $\Back_D^T(z,l)$ and $\HdBack_D^T(z,l)$.}
\label{fig:backarcs}
\end{figure}

We now define the essential concept of backward arcs more formally, which is illustrated in Figure~\ref{fig:backarcs}. This figure shows only the arcs of the out-tree $T$ of $D$, and a set of backward arcs that will be called $\Back_D^T(z,l)$. For arcs that are part of $T$, the arc directions are not shown. The convention in our figures will be that those arcs are directed from left to right.
Let $T$ be an out-tree of $D$, let $l$ be a leaf of $T$ and $z\preceq_T l$. Then
\[
\Back_D^T(z,l)=\{(u,v)\in A(D): (v\prec_T z\preceq_T u\preceq_T l)\}.
\]
Loosely speaking, this is the set of arcs of $D$ that have their heads before $z$, and tails between $z$ and $l$. For our algorithm it is relevant how many different vertices are heads of such arcs, but not out-neighbors of branch vertices, so let
\[
\HdBack_D^T(z,l)=\Hd(\Back_D^T(z,l))\bs \BO(T).
\]
If it is clear what the graphs $D$ and $T$ in question are, the subscript and superscript will be omitted. Informally speaking, we will show that when $|\HdBack_D^T(z,l)|$ is large for some choice of $z$ and $l$, an out-branching with at least $k$ leaves can be found, provided that $D$ contains no useless arcs. On the other hand, when this quantity is small for every choice of $z$ and $l$, a path decomposition of $D$ can be found with small width, which allows us to do dynamic programming.

\section{An FPT Algorithm for Directed Spanning $k$-Leaf}

\subsection{Overview of the Algorithm}
\label{sect:overview}

In this section we give an overview of our FPT algorithm, Algorithm~\ref{alg:FPTSpankleaf}. Details of the different steps are given in the next subsections, and in Section~\ref{sect:algsummary} we combine these to prove the correctness and time complexity of the algorithm.
The main idea is as follows: we consider every possible root $r$, and only consider the arcs that are not useless for this root. In polynomial time, we can construct a 1-optimal out-branching $T$ rooted at $r$. If the number of backward arcs at some point is large, an out-branching with at least $k$ leaves exists. Otherwise, if $T$ itself also has less than $k$ leaves, a path decomposition of bounded width can be found, which enables a dynamic programming procedure.

\begin{algorithm}[h]
    \dontprintsemicolon
    \Input{A digraph $D$ and integer $k$.}
    \BlankLine\;
      
    \For{all $r\in V(D)$ with $R_D(r)=V(D)$}{
        \lnl{removearcs}
        Remove from $D$ all useless arcs for $r$ and obtain $D'$.\;
        \lnl{oneopt}
        Compute a 1-optimal out-branching $T$ of $D'$ with root $r$.\;	
        \lnl{enoughleaves}
	\lIf{$|\LL(T)|\geq k$}{Return(YES).}\;
        \lnl{manybackarcs}
	\lIf{$T$ has a leaf $l$ and vertex $x\prec_T l$ such that $\HdBack_{D'}^T(x,l)\geq 6k^2$}
	   {Return(YES).}\;
	\lnl{pathdecomp}
	Construct a path decomposition of $D'_u$ with width at most $6k^3$.\;
        \lnl{dynprog}
        Do dynamic programming on the path decomposition of $D'_u$.\;
        \lnl{outbranchingfound}
        \lIf{an out-branching with at least $k$ leaves is found}{Return(YES).}
    }
    \lnl{returnno}
    Return(NO)\;
    \caption{An FPT algorithm for {\sc Directed Spanning $k$-Leaf}.}
    \label{alg:FPTSpankleaf}
\end{algorithm}

In Section~\ref{sect:uselessarcs} we will characterize the useless arcs and show how to remove them in polynomial time. In Section~\ref{sect:oneoptoutb} we will deduce properties of 1-optimal out-branchings and show how to find them in polynomial time. In Section~\ref{sect:mainlemma} we prove the key lemma; we prove the correctness of Step~\ref{manybackarcs}. The construction of the path decomposition is treated in Section~\ref{sect:pathdecomp}.

\subsection{Removing useless arcs}
\label{sect:uselessarcs}

In this section, we give a polynomial time algorithm for removing the useless arcs for some give vertex $r$ of digraph $D$.

\begin{proposition}
\label{propo:char_useless}
Let $D$ be a digraph with a vertex $r$ such that $R_D(r)=V(D)$. An arc $(u,v)$ of $D$ is not useless for $r$ if and only if there is a dipath in $D$ starting at $r$ that ends with $(u,v)$.
\end{proposition}
\proof
Clearly, if $(u,v)$ is not useless for $r$ then the out-branching with $r$ as root that contains $(u,v)$ also contains the desired dipath.
Now let $P$ be a dipath starting at $r$ and ending with $(u,v)$. Since $R_D(r)=V(D)$, Proposition~\ref{propo:extendingouttree} shows this dipath can be extended to an out-branching containing $P$.
\QED

We employ this observation to derive our subroutine:

\begin{algorithm}[h]
    \dontprintsemicolon
    \Input{A digraph $D$ and a vertex $r \in V(D)$.}
    \BlankLine\;
      
    \For{all $(u,v)\in A(D)$ where $r \neq u,v$}{
         Test if there is a dipath in $D - v$ from $r$ to $u$.\;
 \lIf{ no such dipath exists}{remove $(u,v)$ from $A(D)$.}
}
\BlankLine\;
\Output{Digraph $D$ with no useless arcs for $r$.}
    \caption{Subroutine to Algorithm~\ref{alg:FPTSpankleaf}, Line~\ref{removearcs}. A polynomial time algorithm for removing useless arcs for vertex $r$.}
    \label{alg:removeuselessarcs}
\end{algorithm}

\subsection{Computing a $1$-optimal out-branching}
\label{sect:oneoptoutb}

Together, the following two lemmata give sufficient and necessary conditions for when a $1$-change leads to an out-branching with more leaves. We use these to design a simple polynomial time algorithm for computing a $1$-optimal out-branching.

\begin{lemma}
\label{obs:onechange_allowed}
Let $T$ be an out-branching of $D$, and let $(u,v)\in A(D)\bs A(T)$. The 1-change for $(u,v)$ gives again an out-branching of $D$ if and only if $v\not\preceq_T u$.
\end{lemma}
\proof
Let $T'$ be the result of a 1-change for $(u,v)$ on an out-branching $T$.
Note that $|A(T')|=|A(T)|$, and that the in-degrees have not changed. Therefore $T'$ is again an out-branching if and only if $T'$ contains no cycles (directed or undirected).

Suppose $T'$ contains an (undirected) cycle $C$. Then this cycle must contain $(u,v)$. Since $v$ again has in-degree 1 in $T'$, the next arc of $C$ must have $v$ as tail. Since all vertices in $T'$ have in-degree at most 1, we can continue the argument like this, and conclude that $C$ is in fact a directed cycle, and thus contains a $(v,u)$-dipath. This dipath is also part of $T$, so $v\preceq_T u$.

To prove the other direction, suppose $v\preceq_T u$, so $T$ contains a $(v,u)$-dipath $P$. This path does not contain $(w,v)$ where $w$ is the unique in-neighbor of $v$, so $P$ is again part of $T'$. It follows that $T'$ contains a (directed) cycle.
This concludes the proof.\QED

\begin{lemma}
\label{obs:onechange_impr}
Let $T$ be an out-branching of $D$, and let $(u,v)\in A(D)\bs A(T)$. The 1-change for $(u,v)$ increases the number of leaves if and only if $u\not\in \LL(T)$ and $v\not\in \BO(T)$.
\end{lemma}
\proof
Adding $(u,v)$ and removing $(w,v)$, where $w$ is the in-neighbor of $v$, increases the out-degree of $u$, and decreases the out-degree of $w$, and changes no other out-degrees. So if $u$ was a leaf before, it loses leaf status. Vertex $w$ becomes a leaf if and only if it was not a branch vertex before. No other vertices gain or lose leaf status. The statement follows.
\QED

Algorithm~\ref{alg:computeoneopt} now shows how to find a 1-optimal out-branching. Correctness follows from the above lemmata, and the algorithm terminates in polynomial time since every iteration increases $|\LL(T)|$. Note that the proof of Proposition~\ref{propo:extendingouttree} gives an easy way to find an initial out-branching $T$ in polynomial time.

\begin{algorithm}[h]
    \dontprintsemicolon
    \Input{A digraph $D$ with no useless arc for  vertex $r \in V(D)$.}
    \BlankLine\;
   Create an out-branching $T$ of $D$ rooted at $r$.\;
    \For{all $(u,v)\in A(D)\setminus A(T)$}{
       \lIf{$v\not\preceq_T u$ {\bf and} $u\not\in \LL(T)$  {\bf and} $v\not\in \BO(T)$ }{Do the $1$-change for $(u,v)$.}
}
\BlankLine\;
\Output{$1$-optimal out-branching $T$ of $D$ rooted at $r$.}
    \caption{Sub-routine to Algorithm~\ref{alg:FPTSpankleaf}, Line~\ref{oneopt}. A polynomial time algorithm for computing a $1$-optimal out-branching.}
    \label{alg:computeoneopt}
\end{algorithm}

\subsection{The Existence of Out-branchings with Many Leaves}
\label{sect:mainlemma}

In this section, we prove the key lemma of our algorithm: we prove the correctness of Step~\ref{manybackarcs}. But first we prove some auxiliary lemmata that we will use in its proof.

\begin{lemma}
\label{obs:changes_root_path}
Let $T$ be an out-branching of $D$ and let $Q$ be a dipath in $D$ that starts at the root $r$ of $T$. Then making all of the 1-changes for every arc in $A(Q)\bs A(T)$ yields again an out-branching of $D$ that contains $Q$.
\end{lemma}
\proof
The proof is by induction over $|A(Q)\bs A(T)|$. If this number is zero, then the statement is obvious.

Now let $(u,v)$ be the first arc of $Q$ that is not part of $T$, and let $Q_1$ be the subpath of $Q$ from $r$ to $u$.
This path contains all vertices $w$ with $w\preceq_T u$, so it follows that $v\not\preceq_T u$.
Making the 1-change for $(u,v)$ then yields again an out-branching (Lemma~\ref{obs:onechange_allowed}). After making this 1-change, the resulting out-branching has one more arc in common with $Q$, and we can use induction.
\QED

\begin{lemma}
\label{obs:ub_BO}
Let $T$ be an out-tree. Then $|\BO(T)|\leq 2|\LL(T)|-2$.
\end{lemma}
\proof
By induction over the number of branch vertices of $T$. If $T$ has no branch vertices, the statement is obvious, since $T$ has at least one leaf.

Otherwise, consider a branch vertex $v$ of $T$ such that there are no other branch vertices in $R_T(v)$.
Let $d_o\geq 2$ denote the out-degree of $v$ in $T$, and let $T'$ be the out-tree obtained from $T$ by deleting all vertices in $R_T(v)$ except $v$ itself. So $v$ is a leaf of $T'$, but $d_o$ leaves of $T$ are deleted, and exactly $d_o$ vertices in $\BO(T)$ are deleted. So $|\LL(T')|=|\LL(T)|-d_o+1$, and $|\BO(T')|=|\BO(T)|-d_o$. By induction, $|\BO(T')|\leq 2|\LL(T')|-2$. Then 
\[
|\BO(T)|=|\BO(T')|+d_o\leq 2|\LL(T')|-2+d_o=
\]
\[
2|\LL(T)|-d_o\leq 2|\LL(T)|-2.
\]
\QED

\begin{figure}[h]
\centering
$\input{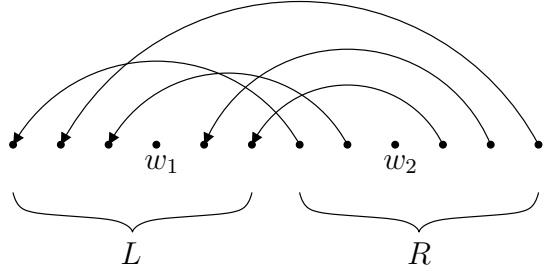}$
\caption{The arcs in $B$ and two elements in $W$.}
\label{fig:nicelyinthemiddle1}
\end{figure}

Figure~\ref{fig:nicelyinthemiddle1} illustrates the next lemma. 
Here $L$, $R$, $\prec$ and $B$ are represented by a graph with vertex set $L\cup R$, drawn from left to right corresponding to $\prec$. The tuples of $B$ are represented by arcs between the corresponding vertices. The set $W=\{w_1,w_2\}$ satisfies the condition from the lemma. Note that for both choices of $w\in W$, there are three arcs $(u,v)\in B$ with $v\prec w\preceq u$.

\begin{lemma}
\label{obs:nicelyinthemiddle}
Let $\preceq$ and $\prec$ be a linear order and corresponding strict linear order on a set $L\cup R$, 
such that $v\in L$ and $u\in R$ imply $v\prec u$.
Let $B$ be a set of at least $2k-1$ 2-tuples $(u,v)$ with $u\in R$ and $v\in L$, and let $W\subseteq L\cup R$ be a set such that for every tuple $(u,v)$, there is a $w\in W$ with $v\prec w\preceq u$. Then there exists a $w\in W$ such that there are at least $k$ tuples $(x,y)\in B$  with $y \prec w \preceq x$. 
\end{lemma}
\proof
We first show that there is a tuple $(u,v)\in B$ such that
\begin{itemize}
\item
there are at least $k$ tuples $(x,y)\in B$ with $y\preceq v$, and 
\item
there are at least $k$ tuples $(x,y)\in B$ with $u\preceq x$.
\end{itemize}
For $v\in L$, let $L(v)$ denote the number of tuples $(x,y)\in B$ with $y\preceq v$ (note that tuples with $y=v$ also count towards $L(v)$ since $\preceq$ is reflexive). For $u\in R$, let $R(u)$ denote the number of tuples $(x,y)$ with $u\preceq x$.
First remove all tuples $(u,v)$ from $B$ with $L(v)\leq k-1$; at most $k-1$ tuples are removed and at least $k$ left. Then remove all tuples $(u,v)$ from $B$ with $R(u)\leq k-1$; at most $k-1$ tuples are removed. At least one tuple is remaining, for which the statement holds. Let $(u,v)$ be this tuple and let $w$ be an element of $W$ with $v \prec w \preceq u$.

We now show that our choice of $(u,v)$ implies that there are at least $k$ tuples $(x,y)\in B$ with $y\prec w\preceq x$.
First suppose $w\in L$. Then for all $(x,y)\in B$, $w\prec x$.
Since we have chosen $v$ such that there are at least $k$ arcs $(x,y)$ in $B$ such that $y\preceq v\prec w$, all of these choices give $y\prec w\preceq x$. The case $w\in R$ is analog.
\QED

Now we are ready to prove the correctness of Step~\ref{manybackarcs} of Algorithm~\ref{alg:FPTSpankleaf}.

\begin{figure}[h]
\centering
$\input{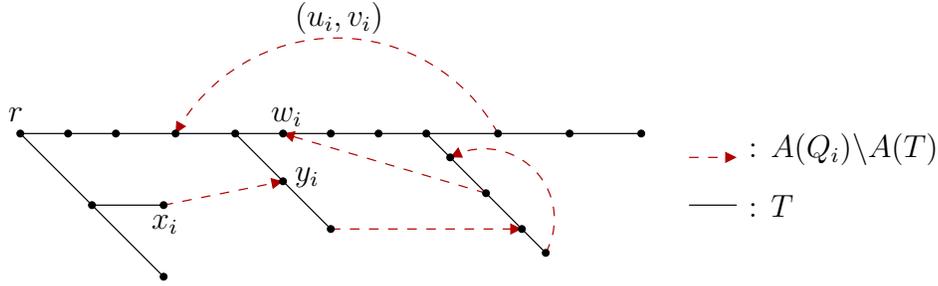}$
\caption{An out-branching $T$, arc $(u_i,v_i)$ and path $Q_i$.}
\label{fig:pathQ}
\end{figure}

\begin{lemma}
Let $D$ be a digraph without useless arcs for $r\in V(D)$, and let $T$ be a 1-optimal out-branching of $D$ rooted at $r$. If there exists a leaf $l$ of $T$ and vertex $z$ with $z\prec_T l$ such that $|\HdBack_D^T(z,l)|\geq 6k^2$, then $D$ has an out-branching with at least $k$ leaves.
\label{lemma:keylemma}
\end{lemma}
\proof
The fact that $|\HdBack(z,l)|\geq 6k^2$ shows that there exists a set of arcs $B\subset A(D)$ such that the following conditions hold:
\begin{enumerate}
\item
$|B|\geq 6k^2$.
\item
\label{cond:linord}
Leaf $l$ is reachable from every vertex in $\Hd(B)\cup \T(B)$.
\item
\label{cond:headsbeforetails}
For $v\in \Hd(B)$ and $u\in \T(B)$, $v\prec_T u$ holds.
\item
\label{cond:disjointheads}
The heads of any two arcs in $B$ are disjoint.
\item
\label{cond:headsnotinBO}
For all $v\in \Hd(B)$, $v\not\in \BO(T)$ holds.
\end{enumerate}
Note that Condition~\ref{cond:linord} implies that $\preceq_T$ is a linear order on $\Hd(B)\cup \T(B)$; this important fact will be used implicitly throughout the proof.
We only have to consider the case that $|\LL(T)|\leq k-1$.
Let $B=\{(u_i,v_i):i=1,\ldots,m\}$.
Since $D$ contains no useless arcs for $r$, for every arc $(u_i,v_i)$ there is a dipath $Q_i$ in $D$ starting at $r$ and ending with $(u_i,v_i)$ (Proposition~\ref{propo:char_useless}). This path contains the following important arcs and vertices: let $x_i$ be the last vertex of $Q_i$ that is not in $R_T(v_i)$.
Let $y_i$ be the next vertex of $Q_i$, so $v_i\prec_T y_i$, and $(x_i,y_i)\in A(Q_i)$.
Let $w_i$ be the first vertex of $Q_i$ with $v_i\preceq_T w_i\preceq_T u_i$. It is possible that $w_i=y_i$ or $w_i=u_i$, or both. 
These definitions are illustrated in Figure~\ref{fig:pathQ}. For arcs that are part of $T$, the arc directions are not shown. The convention in our figures will be that those arcs are directed from left to right.
%%%%%%

\begin{figure}[h]
\centering
$\input{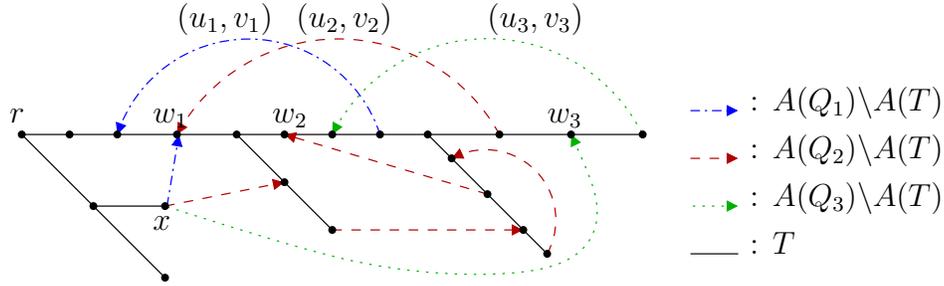}$
\caption{Three arcs in $B'$, the corresponding paths $Q_i$, and vertex $x$.}
\label{fig:threepaths}
\end{figure}

Note that for all $i$, the arc $(x_i,y_i)$ has $y_i\not\preceq_T x_i$ and $(x_i,y_i)\in A(D)\bs A(T)$. So by the 1-optimality of $T$, we know that $x_i\in \LL(T)$ or $y_i\in \BO(T)$ (Lemma~\ref{obs:onechange_allowed}, Lemma~\ref{obs:onechange_impr}). 
Let $B_x=\{(u_i,v_i)\in B: x_i\in \LL(T)\}$, and $B_y=\{(u_i,v_i)\in B: y_i\in \BO(T)\}$. Then$|B_x|+|B_y|\geq 6k^2$, so we may distinguish two cases: $|B_x|\geq 2k^2$ or $|B_y|\geq 4k^2$.\\ 
\\
{\bf Case 1:} $|B_x|\geq 2k^2$.\\
\\ 
Since $|\LL(T)|\leq k-1$, there is a vertex $x$ such that $x=x_i$ for at least $2(k+1)$ choices of $i$. Choose such a vertex $x$, and let $B'\subseteq B_x$ be the set of arcs $(u_i,v_i)$ with $x_i=x$. So $|B'|\geq 2(k+1)$.
%%%%%%
Figure~\ref{fig:threepaths} illustrates these notions. For clarity we have only drawn some arcs in $B'$. 

%%%%%%
By Condition~\ref{cond:headsbeforetails}, we may now use Lemma~\ref{obs:nicelyinthemiddle} for the arc set $B'$, so we can choose 
an arc $(u_i,v_i)\in B'$ such that there are at least $k+1$ arcs $(u_j,v_j)\in B'$ with $v_j\prec_T w_i \preceq_T u_j$. For this choice of $i$, let $B''\subseteq B'$ denote this set of arcs, and let $u=u_i$, $v=v_i$, $y=y_i$ and $w=w_i$.
\begin{figure}[h]
\centering
$\input{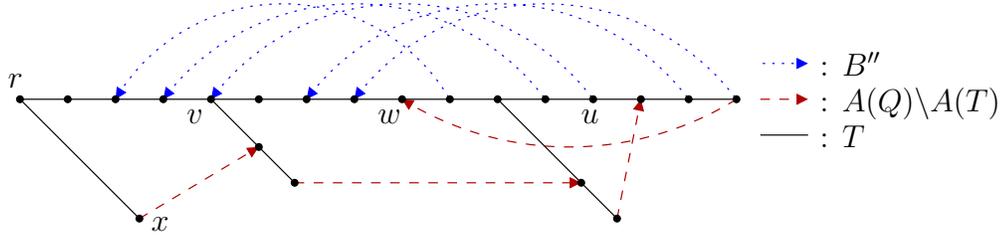}$
\caption{The path $Q$ and arc set $B''$.}
\label{fig:QandBpp}
\end{figure}
%%%%%%
\noindent We construct a dipath $Q$ as follows,
see Figure~\ref{fig:QandBpp}.
\begin{itemize}
\item
Start with the unique path in $T$ from $r$ to $x$. 
\item
Add the subpath of $Q_i$ from $x$ to $w$. 
\end{itemize} 
The first path contains only vertices that are not in $R_T(v)$, and the second path contains only vertices in $R_T(v)$, except $x$, so these paths only share $x$. 
Hence the resulting $Q$ is a dipath from $r$ to $w$.

\begin{figure}[h]
\centering
$\input{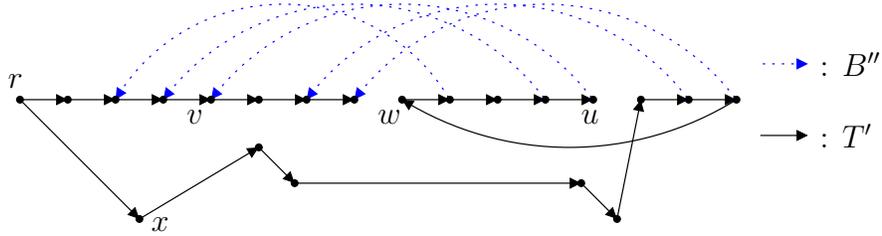}$
\caption{The out-branching $T'$ obtained by making 1-changes for $A(Q)\bs A(T)$.}
\label{fig:TpandBpp}
\end{figure}

Now let $T'$ be obtained by making all 1-changes for arcs in $A(Q)\bs A(T)$,
%%%%%%
see Figure~\ref{fig:TpandBpp}.
%%%%%%
By Lemma~\ref{obs:changes_root_path}, $T'$ is again an out-branching, which contains $Q$.

Now we show that for all arcs $(u_j,v_j)\in B''$, we have $v_j\not\preceq_{T'} u_j$.
Suppose $T'$ contains a dipath $P$ from $v_j$ to $u_j$. Since $A(Q)\subseteq A(T')$ and $Q$ starts at the root of $T'$, if $P$ contains at least one vertex from $Q$ then its first vertex $v_j$ is also in $Q$. But $Q$ started with a path in $T$ from $r$ to $x$, and $v_j\not\preceq_T x$. In addition, for all subsequent vertices $a\in V(Q)$, $a\not\prec_T w$ holds, but $v_j\prec_T w$. 
So $v_j\not\in V(Q)$, and
thus $P$ does not share vertices with $Q$. Then $P$ is also a path in $T$. But the unique path from $v_j$ to $u_j$ in $T$ contains $w\in V(Q)$, a contradiction.

\begin{figure}[h]
\centering
$\input{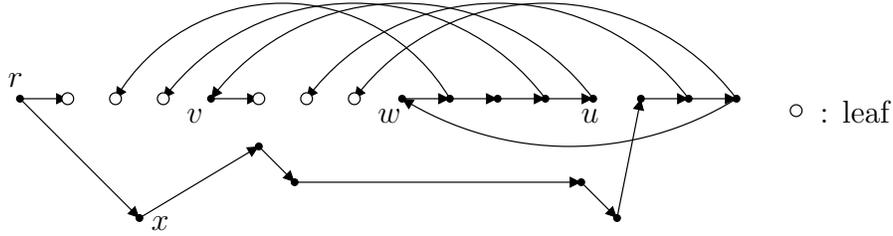}$
\caption{The leafy tree obtained from $T'$ by making 1-changes for $B''$.}
\label{fig:leafytree}
\end{figure}

This shows that if we make 1-changes for all arcs in $B''$, again an out-branching is obtained (Lemma~\ref{obs:onechange_allowed}); note that making one such 1-change does not change the fact that $v_j\not\preceq_{T'} u_j$ for all other $(u_j,v_j)\in B''$. We now argue that this introduces at least $k$ leaves.
Recall that all arcs in $B''$ have disjoint heads (Condition~\ref{cond:disjointheads}).
Let $v_j\in \Hd(B'')$ be the minimum among $\Hd(B'')$ with respect to $\preceq_T$. This is then the only vertex in $\Hd(B'')$ that may have an in-neighbor in $Q$, and thus the only vertex for which the out-degree of its in-neighbor may be different in $T$ and $T'$. So for all of the other vertices in $\Hd(B'')$, the out-degree of their in-neighbor is the same in $T$ and $T'$, so they are again not in $\BO(T')$ (Condition~\ref{cond:headsnotinBO}). Making the 1-changes for these arcs makes their in-neighbors leaves, and thus yields at least $k$ leaves,
%%%%%%
see Figure~\ref{fig:leafytree}.
%%%%%%

\medskip

\noindent
{\bf Case 2:} $|B_y|\geq 4k^2$.\\
\\
Since $|\BO(T)|\leq 2(k-1)$ (Lemma~\ref{obs:ub_BO}), there is a vertex $y$ such that $y=y_i$ for at least $2(k+1)$ choices of $i$. Choose such a vertex $y$, and let $B'\subseteq B_y$ be the set of arcs $(u_i,v_i)$ with $y_i=y$. So $|B'|\geq 2(k+1)$.

 Just as in the first case, it follows with Lemma~\ref{obs:nicelyinthemiddle} that there exists an arc $(u_i,v_i)\in B'$ such that there are at least $k+1$ arcs $(u_j,v_j)\in B'$ with $v_j\prec_T w_i\preceq_T u_j$. Let $B''\subseteq B'$ denote the set of those arcs.
For this choice of $i$, let $u=u_i$, $v=v_i$ and $w=w_i$.
Let $v_j\in \Hd(B'')$ be the unique minimum of $\Hd(B'')$ with respect to $\preceq_T$. Let $x=x_j$, so for all $v_l\in \Hd(B'')$ we have $v_l\not\preceq_T x$. We construct a dipath $Q$ as follows.
\begin{itemize}
\item
Start with the unique path in $T$ from $r$ to $x$. 
\item
Add arc $(x,y)$.
\item
Add the subpath of $Q_i$ from $y$ to $w$.
\end{itemize}
At this point, we have chosen and constructed $Q$, $B''$, $x$, $y$ and $w$ with the same essential properties as in the first case, and the proof can be completed in the same way.\QED

\subsection{Finding a Path Decomposition with Bounded Width}
\label{sect:pathdecomp}

In this section we show how to construct the path decomposition of $D'_u$ in Line~5 of Algorithm~\ref{alg:FPTSpankleaf}, and prove that it has width at most $6k^3$. 
For every $u\in V(T)$, let $h(u)$ denote the {\em height} of $u$, that is, the distance in $T$ from the root $r$ to $u$. 
The path decomposition will consist of bags $X_1,\ldots,X_m$,
where $m$ is the maximum height among all vertices.
For $i\in \{1,\ldots,m\}$, bag $i$ will now contain all vertices at height $i$, all leaves of $T$, all vertices in $\BO(T)$, and all vertices with height less than $i$ that have a neighbor at height at least $i$ which is not in $\LL(T)$ or $\BO(T)$. Formally, $X_i$ is defined as follows. Let $R=(V(T)\bs \LL(T))\bs \BO(T)$
\[
\begin{array}{cl}
X_i = & \{v\in V(T): h(v)=i\}\ \cup\ \BO(T)\ \cup\ \LL(T)\ \cup\\
      & \{v\in V(T): \ \exists u\in R:\ (h(v)<i\leq h(u)) \wedge (uv\in E(D'_u))\}.
\end{array}
\]
\begin{lemma}
\label{lem:ispathdecomp}
$(X_1,\ldots,X_m)$ as constructed above is a path decomposition of $D'_u$.
\end{lemma}
\proof
Every vertex $u\in V(D'_u)$ is included in at least one bag, namely $X_{h(u)}$ if $h(u)\geq 1$. The root of $T$ is included in $X_1$. 
Now we show that for every edge $uv\in E(D'_u)$ there is a bag containing both $u$ and $v$. If one of its end vertices, say $v$, is in $\BO(T)$ or in $\LL(T)$, then $u,v\in X_{h(u)}$. Otherwise, suppose w.l.o.g. that $h(v)\leq h(u)$, then we have $u,v\in X_{h(u)}$. 
For condition three, note that if $h(u)=i$ and $u\not\in \BO(T)\cup\LL(T)$, then $u$ does not appear in any bag $X_l$ with $l<i$. Suppose $v$ is the vertex at maximum height $j$ in $T$ such that there is an arc $uv\in E(D'_u)$. Then $w$ is inside every bag $X_i,\ldots,X_j$ and not in any bag $X_l$ with $l>j$. Hence, the bags containing $u$ are consecutive.
\QED

\begin{lemma}
\label{lem:boundedwidth}
Let $T$ be a 1-optimal out-branching of a digraph $D$ 
with $|\LL(T)|\leq k-1$. If for all leaves $l$ of $T$ and vertices $x$ with $x\preceq_T l$ it holds that $|\HdBack_D^T(x,l)|<6k^2$, then the path decomposition $(X_1,\ldots,X_m)$ as constructed above has width at most $6k^3$.
\end{lemma}
\proof
Consider a set $X_i$. 
We will show that for every $v\in X_i$ one of the following statements holds:
\begin{itemize}
\item
$i-1\leq h(v)\leq i$, or
\item
$v\in \LL(T)$, or
\item
$v\in \BO(T)$, or
\item
There is an $l\in \LL(T)$ such that for the unique vertex $x\preceq_T l$ with $h(x)=i$, $v\in \HdBack(x,l)$ holds.
\end{itemize}
Combined with upper bounds on the sizes of these sets, the statement will follow.
Suppose $h(v)\not=i$, $h(v)\not=i-1$, $v\not\in\LL(T)$ and $v\not\in\BO(T)$. By the definition of $X_i$, $h(v)<i$ and $v$ has a neighbor $u$ with $u\not\in \LL(T)$, $u\not\in \BO(T)$ and $h(u)\geq i$. Then $h(v)\leq i-2$ and $h(u)\geq i$, which shows that the arc between those vertices is not in $T$.
Since $T$ is 1-optimal and both $u$ and $v$ are not in $\LL(T)$ or $\BO(T)$, Lemma~\ref{obs:onechange_allowed} and Lemma~\ref{obs:onechange_impr} show that either $v\prec_T u$ or $u\prec_T v$. The latter is not possible since $h(u)>h(v)$, so we have $v\prec_T u$. Now let $l$ be a leaf of $T$ that is reachable from $u$, and let $x$ be the unique vertex with $x\preceq_T l$ and $h(x)=i$. Then we have $v\prec_T x\preceq_T u\preceq_T l$, so $v\in \HdBack(x,l)$. 
Note that since this choice of $x$ is unique for every $l$, we have proved that $v$ is part of one out of at most $|\LL(T)|$ possible sets $\HdBack(x,l)$; one for every choice of $l$.

Since $|\LL(T)|\leq k-1$, we have the following upper bounds: $|\{v\in V(T):i-1\leq h(v)\leq i\}|\leq 2k-2$, $|\BO(T)|\leq 2k-4$ (Lemma~\ref{obs:ub_BO}), and for every $l\in \LL(T)$ and $x\prec_T l$, $\HdBack(i,l)<6k^2$. It follows that 
\[
|X_i|< (k-1) + (2k-2) + (2k-4) + (k-1)6k^2 < 6k^3.
\]
\QED

\subsection{Summary of the Algorithm}
\label{sect:algsummary}

The only step of Algorithm~\ref{alg:FPTSpankleaf} that we have not explained in detail is Step~\ref{dynprog}, the dynamic programming step. 
When a tree or path decomposition is given of the underlying undirected graph of $D'$, standard dynamic programming methods can be used to decide whether $D'$ has an out-branching with at least $k$ leaves. The time complexity of such a procedure is $2^{O(w\log k)}\cdot n$, where $n=|V(D)|$ and $w$ is the width of the path decomposition. For details, see e.g.~\cite{AlonFGKS07}.

\begin{thm}
For any digraph $D$ with $n=|V(D)|$, Algorithm~\ref{alg:FPTSpankleaf} solves the problem {\sc Directed Spanning $k$-Leaf} in time $2^{O(k^3 \log k)} \cdot n^{O(1)}$.
\end{thm}
\proof
We first prove that Algorithm~\ref{alg:FPTSpankleaf} returns the correct answer in every case.
Step~\ref{enoughleaves} and~\ref{outbranchingfound} are obviously correct.
The correctness of Step~\ref{manybackarcs} is given by Lemma~\ref{lemma:keylemma}.
In Section~\ref{sect:pathdecomp} it is shown that if YES is not returned in Step~\ref{manybackarcs}, then in Step~\ref{pathdecomp} indeed a path decomposition with width at most $6k^3$ can be constructed.

Now we prove the correctness of Step~\ref{returnno}, hence we prove that if an out-branching with at least $k$ leaves exists in $D$, it will be found during some iteration of the for-loop. Suppose $T$ is an out-branching of $D$ with at least $k$ leaves and root $r$. 
Clearly, $R_D(r)=V(D)$, so consider the iteration of the for-loop where this $r$ is considered. By definition of useless arcs, $T$ is also an out-branching of $D'$. The dynamic programming procedure considers all possibilities, so in Step~\ref{outbranchingfound} YES is returned, if not before in Step~\ref{enoughleaves} or~\ref{manybackarcs}.

Finally we consider the time complexity of Algorithm~\ref{alg:FPTSpankleaf}. We have shown that every step of the algorithm can be done in time polynomial in $n$, except Step~\ref{dynprog}, which takes time $2^{O(k^3 \log k)} \cdot n$, since the width of the path decomposition is bounded by $6k^3$ (Lemma~\ref{lem:boundedwidth}). Steps~\ref{removearcs}--\ref{outbranchingfound} are repeated at most $n$ times (for every possible choice of the root), so in total the complexity becomes $2^{O(k^3 \log k)} \cdot n^{O(1)}$.\QED

Note that Algorithm~\ref{alg:FPTSpankleaf} can be made into a constructive FPT algorithm, since the proof of Lemma~\ref{lemma:keylemma} can be turned into a polynomial time algorithm that constructs an out-branching.

\bibliographystyle{siam}
\bibliography{digraph2}

\begin{thebibliography}{10}

\bibitem{AlonFGKS07II}
{\sc N.~Alon, F.~V. Fomin, G.~Gutin, M.~Krivelevich, and S.~Saurabh}, {\em
  Better algorithms and bounds for directed maximum leaf problems}.
\newblock Preprint: http://www.arxiv.org/pdf/0707.1095, 2007.

\bibitem{AlonFGKS07}
\leavevmode\vrule height 2pt depth -1.6pt width 23pt, {\em Parameterized
  algorithms for directed maximum leaf problems}, in Proceedings of the 34th
  International Colloquium on Automata, Languages and Programming (ICALP 2007),
  vol.~4596 of LNCS, Springer, 2007, pp.~352--362.

\bibitem{AlonYZ95}
{\sc N.~Alon, R.~Yuster, and U.~Zwick}, {\em Color-coding}, J. ACM, 42 (1995),
  pp.~844--856.

\bibitem{BangJG00}
{\sc J.~Bang-Jensen and G.~Gutin}, {\em Digraphs: Theory, Algorithms and
  Applications}, Springer-Verlag, 2000.

\bibitem{BonsmaBW03}
{\sc P.~S. Bonsma, T.~Br{\"u}ggemann, and G.~J. Woeginger}, {\em A faster {FPT}
  algorithm for finding spanning trees with many leaves}, in Proceedings of the
  28th International Symposium on the Mathematical Foundations of Computer
  Science (MFCS 2003), vol.~2747 of LNCS, Springer, 2003, pp.~259--268.

\bibitem{bonsmazickfeld}
{\sc P.~S. Bonsma and F.~Zickfeld}, {\em Spanning trees with many leaves in
  graphs without diamonds and blossoms}.
\newblock Accepted for LATIN 2008. Preprint: http://arxiv.org/pdf/0707.2760,
  2007.

\bibitem{Cesati}
{\sc M.~Cesati}, {\em Compendium of parametrized problems}.
\newblock http://bravo.ce.uniroma2.it/home/cesati/research/compendium.pdf,
  2006.

\bibitem{DemaineGMS07}
{\sc E.~Demaine, G.~Gutin, D.~Marx, and U.~Stege}, {\em Open problems from
  dagstuhl seminar 07281: Structure theory and {FPT} algorithmics for graphs,
  digraphs and hypergraphs}, 2007.
\newblock manuscript, http://erikdemaine.org/papers/DagstuhlFPT2007Open/.

\bibitem{Diestel}
{\sc R.~Diestel}, {\em Graph Theory}, no.~173 in Graduate Texts in Mathematics,
  Springer-Verlag, New York, 1997.

\bibitem{DingJS01}
{\sc G.~Ding, T.~Johnson, and P.~Seymour}, {\em Spanning trees with many
  leaves}, Journal of Graph Theory, 37 (2001), pp.~189--197.

\bibitem{FellowsMRS00}
{\sc M.~R. Fellows, C.~McCartin, F.~A. Rosamond, and U.~Stege}, {\em
  Coordinatized kernels and catalytic reductions: An improved {FPT} algorithm
  for max leaf spanning tree and other problems}, in Proceedings of the 20th
  International Conference on the Foundations of Software Technology and
  Theoretical Computer Science (FSTTCS 00), vol.~1974 of LNCS, Springer, 2000,
  pp.~240--251.

\bibitem{FominGK06}
{\sc F.~V. Fomin, F.~Grandoni, and D.~Kratsch}, {\em Solving connected
  dominating set faster than $2^n$}, in Proceedings of the 26th International
  Conference on the Foundations of Software Technology and Theoretical Computer
  Science (FSTTCS 06), vol.~4337 of LNCS, Springer, 2006, pp.~152--163.

\bibitem{GalbiatiMM97}
{\sc G.~Galbiati, A.~Morzenti, and F.~Maffioli}, {\em On the approximability of
  some maximum spanning tree problems}, Theor. Comput. Sci., 181 (1997),
  pp.~107--118.

\bibitem{HuangH89}
{\sc N.-F. Huang and T.-H. Huang}, {\em On the complexity of some arborescences
  finding problems on a multihop radio network}, BIT Numerical Mathematics, 29
  (1989), pp.~212--216.

\bibitem{KW91}
{\sc D.~J. Kleitman and D.~B. West}, {\em Spanning trees with many leaves},
  SIAM Journal on Discrete Mathematics, 4 (1991), pp.~99--106.

\bibitem{Liang02}
{\sc W.~Liang}, {\em Constructing minimum-energy broadcast trees in wireless ad
  hoc networks}, in Proceedings of the 3rd ACM international symposium on
  Mobile ad hoc networking and computing, 2002, pp.~112--122.

\bibitem{LuR98}
{\sc H.-I. Lu and R.~Ravi}, {\em Approximating maximum leaf spanning trees in
  almost linear time}, J. Algorithms, 29 (1998), pp.~132--141.

\bibitem{Solis-Oba98}
{\sc R.~Solis-Oba}, {\em 2-approximation algorithm for finding a spanning tree
  with maximum number of leaves}, in Proceedings of the 6th Annual European
  Symposium on Algorithms (ESA 98), vol.~1461 of LNCS, Springer, 1998,
  pp.~441--452.

\bibitem{TWL07}
{\sc M.~Thai, F.~Wang, D.~Liu, S.~Zhu, and D.~Du}, {\em Connected dominating
  sets in wireless networks with different transmission ranges}, IEEE
  transactions on mobile computing, 6 (2007), pp.~721--730.

\end{thebibliography}

\end{document}